\documentclass[superscriptaddress,twocolumn,showkeys,amsmath,amssymb,pre,showpacs]{revtex4-1}

\usepackage{bm,bbm}
\usepackage{graphicx}
\usepackage[table]{xcolor}

\newcommand{\eps}{{\varepsilon}}

\newcommand{\rmd}{{\rm d}}
\newcommand{\rme}{{\rm e}}
\newcommand{\rmi}{{\rm i}}
   
\newcommand{\la}{\langle}
\newcommand{\ra}{\rangle}

\begin{document}
\title{Mode resolved travel time statistics for elastic rays in 
  three-dimensional billiards}
\author{A. Ortega} \affiliation{Instituto de F\'isica, Universidad Nacional
  Aut\' onoma de M\' exico, M\' exico D. F. 01000, M\' exico}
\author{K. Stringlo}
\author{T. Gorin} \affiliation{Departamento de F\'isica, Universidad de 
  Guadalajara, Guadalajara 44840, Jal\'isco, M\'exico}
\date{\today}


\begin{abstract}
We consider the ray limit of propagating ultrasound waves in three-dimensional 
bodies made from an homogeneous, isotropic, elastic material. Using a Monte 
Carlo approach, we simulate the propagation and proliferation of elastic rays
using realistic angle dependent reflection coefficients, taking into account 
mode conversion and ray-splitting. For a few simple geometries, we analyse
the long time equilibrium distribution focussing on the energy ratio between 
compressional and shear waves. Finally, we study the travel time statistics, 
{\it i.e.} the distribution of the amount of time a given trajectory spends
as a compressional wave, as compared to the total travel time. These results 
are intimately related to recent elastodynamics experiments on Coda wave
interferometry by Lobkis and Weaver [Phys. Rev. E 78, 066212 (2008)].
\end{abstract}


\maketitle

\section{Introduction}

Elastic rays are the fundamental building block of the theory of Coda wave 
interferometry~\cite{Sni06} which over recent years has been developed into a 
well established method for the analysis of seismological 
data~\cite{Courtland08}, among others~\cite{Larose06}.
To the best of our knowledge, only little is known about elastic rays as the 
particle limit of the elastic wave equation, even though it is precisely that 
limit the theory in Ref.~\cite{Sni06} relies upon. The main difficulty with 
elastic rays is due to the phenomenon of ``ray-splitting'' which causes the 
dynamics to become effectively statistical in nature~\cite{KohBlu98}.

There is a close analogy in elastic rays and Coda
wave interferometry on the one hand and the orbits of classically chaotic
quantum systems and semiclassical theory in the diagonal 
approximation~\cite{Berry85,JacBee01,CerTom02} on the other. This analogy is
crucial also for measurable quantities such as the distortion (introduced
in Ref.~\cite{LobWea03}) and scattering 
fidelity~\cite{SGSS05,SSGS05,GSW06,GPSZ06}. In Ref.~\cite{LobWea03} Lobkis and 
Weaver started a series of experiments with reverberant ultrasound in 
three-dimensional Aluminum samples. These experiments could be described in
terms of the Coda wave interferometry (based on the ray picture of a diffuse
wave field), but also in terms of scattering fidelity and random matrix 
theory~\cite{GSW06,LobWea08}. So far the focus has been on the form of the 
fidelity decay~\cite{LobWea08}, but not on the overall decay time of the 
fidelity signal. In order to explain their results
Lobkis and Weaver assumed that after a certain transient time, the 
elastic wave field settles on a equilibrium state, where the energy is 
distributed homogeneously and isotropically over the whole Aluminum sample,
with relative energy shares in shear- and compressional waves corresponding to 
the equipartition ratio~\cite{Wea82,LobWea03}. 


In this paper, we present numerical simulations of the propagation of elastic
rays in finite three-dimensional bodies. Due to ray-splitting an elastic ray 
spreads into an exponentially increasing number of different branches. This 
makes it very difficult to simulate ray-dynamics over long times. One of our
main achievements consists therefore in the development of an efficient 
algorithm which applies a Monte Carlo approach to deal with these 
difficulties. The bodies employed are similar to the ones used by Lobkis and 
Weaver but not identical. Still our simulations allow to verify certain 
assumptions made about the wave field and to analyse the effects of eventual 
violations. 


Together with the introduction, this paper is divided into 6 sections. The 
following Sec.~\ref{R} defines our concept of an elastic ray. Sec.~\ref{M} 
describes the random mode conversion model introduced in Ref.~\cite{LobWea03}.
Our main results are described in Sec.~\ref{N}, and their relation to 
experiments is discussed in Sec.~\ref{D}. We conclude the paper with 
Sec.~\ref{C}.

\section{\label{R} Elastic rays}

Rays are a well known concept from geometric optics, where they are used to 
approximate the propagation of electromagnetic waves in the limit where the 
wavelength is small compared to the typical dimension $L$ of the system. A 
single ray stands for a transversally concentrated electromagnetic wavepacket 
which moves through an optical system. In the propagation direction, the 
wavepacket may also be localized, but that need not be so. Without 
localization one arrives at a stationary situation, where wave energy 
continuously flows along the ray path.

In the present case, we assume that the wavepacket is finite, even in the
longitudinal direction. It therefore contains a finite amount of energy $Q$.
Since we neglect any form of energy dissipation, that amount of energy is
conserved for all times. However, due to mode conversion and ray-splitting,
the amount of energy $Q$ is shared among an increasing number of branches of 
the elastic ray. 

Elastic waves in a homogeneous and isotropic medium come in two different 
modes: $P$-waves (compressional waves) where the medium undergoes harmonic 
displacements in the longitudinal (propagation) direction, and $S$-waves, 
where the medium undergoes harmonic displacements in the transversal direction 
(shear waves). Both wave forms have different propagation speeds, $c_d$ and 
$c_s$ respectively. For the Aluminum samples studied 
in~\cite{LobWea03,GSW06,LobWea08}:
\begin{equation}
c_d = 635\, {\rm cm/ms}\; , \qquad
c_s = 315\, {\rm cm/ms}\; . 
\label{wavespeeds}\end{equation}
As we shall see below, both wave forms convert according to specific rules
into one-another when the ray is reflected at the free surfaces of the body.
In the classical ray picture this is handled by providing the 
elastic ray with an additional internal degree of freedom, as will be explained
in detail below, in Sec.~\ref{RC}.

One should be well aware of the fact that the ray picture, when applied to 
propagating elastic waves in a finite solid, breaks down very soon. 
The transversal size of a localized wavepacket increases rapidly
in time such that after a few reflections, it reaches the extention of the 
whole body. In quantum mechanics this time scale would be called the 
Ehrenfest time~\cite{SchoJac05}. However, this
does not mean that the ray picture becomes useless when longer times are 
involved. In the field of quantum chaos it is well established that one can 
construct semiclassical approximations, on nothing else than these 
classical trajectories~\cite{Gutzwiller90}. These approximations 
may remain valid for much longer times -- in the case of chaotic two
degree-of-freedom systems up to times of the order of the Heisenberg 
time~\cite{Kap02}. Ultimately, the present work might help to pave the way
towards a similar semiclassical theory for elastodynamic systems.

For the numerical simulations, we use geometries (rectangular 
block, tetrahedron) with long straight edges, because similar bodies have been
used in the experiments by Lobkis and Weaver and also because of their 
simplicity. These bodies however have the disadvantage that diffraction on the
edges and corners may have considerable effects~\cite{BPS00,HHH00}. The 
inclusion of diffractive orbits will be left to a future work.

\subsection{\label{RR} Reflection coefficients}

In general, rays are of a hybrid nature. From a macroscopic perspective, the 
ray has negligible width and is regarded as a one-dimensional object, while,
from a microscopic perspective, the ray is seen as a plane wave which allows
to use relatively simple rules to calculate its behavior under reflections.
Since the system is assumed to be macroscopic in size, the shape of the 
reflecting surface does not really matter. It is always approximated locally 
as a plane surface. For our purposes it is therefore sufficient to consider 
the reflection laws of plane waves at plane surfaces. For simplicity, we shall 
also assume that all reflecting surfaces are free surfaces.

\begin{figure}[t]
\includegraphics[width=0.5\textwidth]{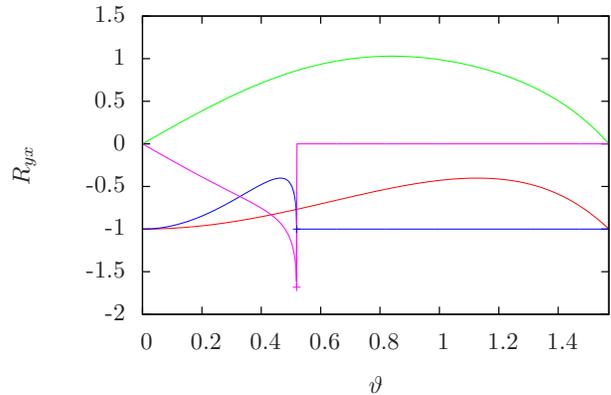}
\caption{The different reflection coefficients as a function of the angle of
incidence $\vartheta$. $R_{PP}$ (red line), $R_{SP}$ (green line),
$R_{SS}$ (blue line), $R_{PS}$ (pink line). Note, $R_{yx}$ means
incident $x$-wave and reflected $y$-wave.}
\label{f:ReflCoefs}\end{figure}

Following~\cite{Ach73}, when a plane elastic wave hits a free surface, we 
first define a reflection plane. This is the plane spanned by the surface 
normal and the propagation direction of the ray. An incident $P$-wave splits 
into an out-going $P$-wave and an out-going $S$-wave with polarization vector 
lying in the plane of reflection. The reflection amplitudes for both waves are 
given by:
\begin{align}
R_{PP} &= \frac{\sin(2\vartheta)\, \sin(2\Theta) - \kappa^2\, \cos^2(2\Theta)}
   {\sin(2\vartheta)\, \sin(2\Theta) + \kappa^2\, \cos^2(2\Theta)}\; , \quad
   \kappa = \frac{\sin\vartheta}{\sin\Theta}
\label{RPP}\\
R_{SP} &= \frac{2\kappa\, \sin(2\vartheta)\, \cos(2\Theta)}
   {\sin(2\vartheta)\, \sin(2\Theta) + \kappa^2\, \cos^2(2\Theta)} \; ,
\end{align}
where $\kappa= c_d/c_s > 1$ such that the exit angle $\Theta$ is always 
smaller than the entrance angle $\vartheta$. These angles are always measured
with respect to the normal of the surface. The reflection amplitudes $R_{PP}$ 
and $R_{SP}$ determine the amplitudes of the two out-going waves. 

The case of an incident $S$-wave is more complicated. Before
considering the reflection itself, we have to decompose the wave into one
component with polarization in the reflection plane (SV-wave) and another
with polarization perpendicular to it (SH-wave). For the SV-wave we then have 
similar reflection coefficients as for the $P$-wave:
\begin{align}
R_{SS} &= 
   \frac{\sin(2\vartheta)\, \sin(2\Theta') - \kappa^2\, \cos^2(2\vartheta)}
   {\sin(2\vartheta)\, \sin(2\Theta') + \kappa^2\, \cos^2(2\vartheta)}\; , 
\quad \kappa = \frac{\sin\Theta'}{\sin\vartheta} \\
R_{PS} &= \frac{-\, \kappa\, \sin(4\vartheta)}
   {\sin(2\vartheta)\, \sin(2\Theta') + \kappa^2\, \cos^2(2\vartheta)} \; ,
\end{align}
where now $\Theta' > \vartheta$ such that these equations only apply as long as
$\kappa\sin\vartheta < 1$. This introduces the critical angle (of incidence)
$\vartheta_{\rm cr}= \arcsin(\kappa^{-1})$. However, for SV-waves incident at 
larger angles, the reflected P-wave becomes a surface wave with negligible 
contribution to the wave field, while the reflection amplitude for the SV-wave 
becomes
\begin{equation}
R_{SS} = -\; 
   \frac{\cos^2(2\vartheta) - 2\rmi\, \beta\, \sin\vartheta\, \sin(2\vartheta)}
   {\cos^2(2\vartheta) + 2\rmi\, \beta\, \sin\vartheta\, \sin(2\vartheta)} \; ,
\label{RSSp}\end{equation}
where $\beta = \sqrt{\sin^2\vartheta - \kappa^{-2}}$. The absolute value 
squared of $R_{SS}$ is one in that case, which means that all the energy is 
transferred to the reflecting $S$-wave.

SH-waves, i.e. shear waves with polarization direction perpendicular to the 
reflection plane, cannot convert to $P$-waves. They are 
reflected according to the standard law where the reflection angle is equal to
the angle of incidence.

\subsection{\label{RC} Classical ray limit}

In order to define the classical ray limit, we consider a localized wavepacket
of total energy $Q$. At any moment in time this wavepacket has a well defined
position $\vec r$ and propagation direction $\vec\rme_v$. With respect to its
momentary position $\vec r$, the wavepacket has a certain extension along the 
propagation direction and perpendicular to it. Typically, we would imagine a
cigar-shaped wavepacket, oriented along the propagation direction. In the 
classical limit, we ignore the transversal extension of the wavepacket and 
obtain thereby a one dimensional object. For the studies to follow, the 
wavepacket's extension along the propagation direction does not matter. Since 
elastic waves propagate in two different modes, the classical description must 
be extended by some internal variables: We need one binary variable to specify 
the mode of the wavepacket which may be longitudinal ($P$) or transversal 
($S$). Moreover, if the wavepacket is in transversal mode, we need to record 
the polarization direction by a unit vector $\vec\rme_p$, perpendicular to the 
propagation direction.

The classical description of elastic waves runs into difficulties when it 
comes to reflections. The problem is ray splitting~\cite{KohBlu98}, which will
be treated as follows: We assume that in the vicinity of the wavepacket center,
an elastic ray may be described as a plane wave. For that plane wave, the
reflection laws of the previous section~\ref{RR} apply. As explained there, the
reflected wave will in general be split into an $S$-mode and a $P$-mode branch, 
propagating into different directions. Accordingly, the reflected ray splits in 
two, where the local intensities and polarization directions in the wavepacket 
center may be calculated from the reflection coefficients of the corresponding 
plane waves. Finally, purely geometric considerations allow to determine the 
widths of those branches. In that way, we obtain the complete information about 
the two reflected branches of the incident ray. Note that an incident $S$-mode 
ray must be considered as a linear combination of a $SV$-wave (polarization 
direction in the reflection plane) and a $SH$-wave (polarization direction 
perpendicular to the reflection plane). Hence, the $P$-wave branch of the 
reflected ray is obtained solely from the $SV$ component, wheras the $S$-wave 
branch is given as a superposition of the reflected $SH$-wave and the $S$-wave 
branch of the $SV$-component. These relations explain why it is necessary to 
keep track of the polarization direction of the $S$-mode rays.

For the classical description of the secondary rays, we need neither the 
wavepacket-widths nor the local intensities, however, we do need the total 
energy share of each branch. In principle, the calculation of the total energy
share is rather involved since it requires the integration of the energy flux
across a surface perpendicular to the propagation direction, taking into 
account the variation of the wavepacket intensity. Fortunately, for the 
reflected branch without mode conversion, neither the wavepacket's intensity 
profile nor its propagation speed change. As a consecuence, the 
total energy contained in that branch is simply proportional to the intensity
calculated from the corresponding reflection coefficients:
\begin{itemize}
\item An incident $P$-wave carrying the energy $Q$ will transfer the 
energy $R_{PP}^2\, Q$ to the reflected $P$-mode branch.
\item An incident $S$-wave of the same energy whose polarization direction 
makes an angle $\theta$ with the reflection plane must be split into the 
$SV$-component of energy $Q_V= Q\, \cos^2\theta$ and the $SH$-component of 
energy $Q_H= Q\, \sin^2\theta$. The total energy of the reflected $S$-mode 
branch then is $R_{SS}^2\, Q_V + Q_H$.
\end{itemize}
Energy conservation now implies that the respective mode-converted branch must 
carry the remaining energy. That this is indeed the case, is shown exemplarily 
for an incident $P$-mode ray in the appendix.

Returning to the propagation of the elastic ray, we note that subsequent 
reflections lead to further subdivisions into an exponentially increasing 
number of branches. Topologically, an elastic ray may be considered as a tree, 
where the initial energy is transported from the trunk towards the outer 
(higher order) branches. Even for modest travel times, it is soon impossible to 
keep track of all the branches of the elastic ray. 

We therefore employ a Monte Carlo method to sample implicitely only over those 
branches which carry the largest amount of energy. For that purpose we 
translate energy share into probability. Thus, we replace the deterministic 
energy distribution model by a statistical model, where we define an ensemble 
of rays which all start in the same initial state corresponding to a wave 
packet with energy $Q_0$. At each reflection, we choose at random, whether the 
ray follows one branch or the other. The probability with which one option or
the other is chosen, are simply given by the relative energy shares calculated 
from the corresponding reflection coefficients. 
At the end, the probability to travel through a certain
higher order branch is given by the product of probabilities for the choices
made along the history of the given ray. This probability agrees with the
energy ratio between the total energy of the wavepacket in that particular 
branch and the energy of the initial wavepacket.
Thus, in a numerical simulation, the energy share of a higher order branch can 
be estimated from the number of members in the ensemble which terminate in that 
given branch. 

Our ensemble of rays may as well start in different initial 
states also chosen at random. That is the case, when we intend to calculate the
evolution of a statistical ensemble of initial conditions. In what follows, we
will consider the following three different types of initial conditions:
\begin{itemize}
\item[{(i)}] {\em Deterministic} initial conditions, where we completely 
specify one particular ray, starting at a certain point, in a specific 
direction, either in $S$- or $P$-mode, and in case with a specific 
polarization. 
\item[{(ii)}] {\em Surface} initial conditions, where we start $P$-waves at a 
particular point on the surface of the body, while the direction of the ray is 
chosen at random within the half sphere pointing into the body.
\item[{(iii)}] {\em Homogeneous} initial conditions, where we start rays 
at random positions inside the body, with random directions (isotropic on the 
whole unit sphere) in $S$- or $P$-mode with probabilities 
chosen according to the equipartition ratio in Eq.~(\ref{theoR}) and in case 
random polarization direction.
\end{itemize}

\section{\label{M} Random mode conversion model}

In Ref.~\cite{LobWea03} the authors introduce a simple model, ignoring any 
geometric effects, where elastic rays undergo random and statistically 
independent mode conversions according to the conversion rates $\alpha$ 
($S$-to-$P$) and $\beta$ ($P$-to-$S$). Here, the mean free time between two conversions is given by $\alpha^{-1}$ (for the 
$S$-mode segments) and $\beta^{-1}$ (for the $P$-mode segments), respectively. 
As an additional ingredient, the authors invoke the equipartition principle,
which states that energy should be distributed equally among the different 
degrees of freedom. Taking into account the different wave speeds for $P$- and
$S$-mode waves, the ratio of the energy share between $S$- and $P$-mode waves
is~\cite{Wea82}
\begin{equation}
R= 2\; \left(\frac{c_d}{c_s}\right)^3\; .
\label{theoR}\end{equation}
This number also determines the ratio between the conversion rates,
since the equilibrium condition of the corresponding rate equation demands
that on average, the number of elastic rays in $P$-mode and $S$-mode are
related by
\begin{equation}
\frac{\la N_S\ra}{\la N_P\ra} = \frac{\beta}{\alpha} = R \; .
\end{equation}
One last piece of information is necessary in order to be able to estimate the
values of the conversion rates. The authors obtain this from an analogy
to room acoustics~\cite{Pierce1981} and a detailed calculation of the mode 
conversion 
probability during individual reflections~\cite{Graff1991}. The result is:
\begin{equation}
\beta = 0.59\; c_d\; \frac{S}{4V}\; .
\label{betatheo}\end{equation}
In order to explain the experiments performed on the various Aluminum samples,
one requires the statistics of $t_P$, the accumulated amount of time a given 
ray of duration $t$ spends in $P$-mode. As an estimate for the
mean $\la t_P\ra$ and its variance ${\rm var}(t_P)$ the authors find
\begin{equation}
\la t_P\ra = \frac{t}{R+1} \; , \qquad
{\rm var}(t_P) = \frac{2t}{\beta}\; \frac{R^2}{(1+R)^3}\; .
\label{MeantpVartp}\end{equation}
The first equation is easily explained on the basis of ergodicity. A more 
involved calculation is required for the second.

\begin{figure}
\includegraphics[width=0.7\textwidth]{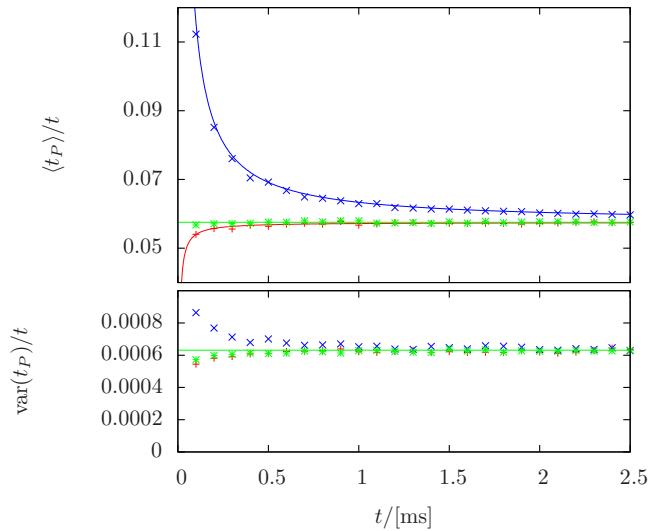}
\caption{Mean and Variance of the $P$-mode travel time for the Lobkis' and
Weavers's random conversion model. The $P$-to-$S$-conversion rate was chosen
as $\beta= 162$. The different colors refer to different initial conditions:
starting in $S$-mode (red), starting in $P$-mode (blue), starting randomly in 
$S$- or $P$-mode according to the equipartition ratio 
(green). 
In the case of the equipartitioned initial condition, the solid green line 
shows the value of the theoretical expectation, Eq.~(\ref{MeantpVartp}).
Other solid lines (red and blue) show simple rational best fit functions to 
guide the eye.}
\label{f:tp-rcm}\end{figure}

For Fig.~\ref{f:tp-rcm} we performed numerical simulations for the random mode 
conversion model described above. We use random numbers with exponential 
probability densities to generate sequences of alternating $P$- and $S$-mode
time segments forming a random realisation of an elastic ray. With the total
time $t$ being fixed we can than measure the average $P$-mode occupation time
$\la t_P\ra$ as well as its variance. These quantities are shown in 
Fig.~\ref{f:tp-rcm} as a function of $t$, for three different cases: When the
trajectory starts with a $P$-mode segment, with a $S$-mode segment, or by
choosing $P$-mode or $S$-mode at random acording to the equipartition ratio.
We find that in all cases, $\la t_P\ra$ and ${\rm var}(t_P)$ quickly converge
to the theoretical values as $t$ becomes sufficiently large.

\section{\label{N} Numerical results}

\begin{figure}
\includegraphics[width=0.4\textwidth]{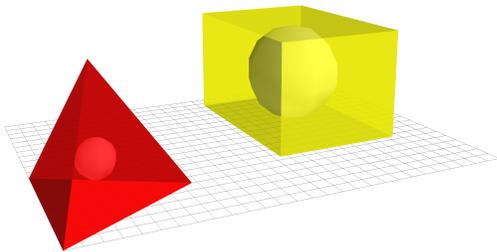}
\caption{On left a regular tetrahedron with length of 10 cm with
inner sphere radius 1.2 cm on right a rectangular block with length of
9 cm width of 13 cm and height of 7.6 cm with inner sphere radius 3.5 cm.}
%
\label{f:geometries}\end{figure}

In our simulations we use bodies of two different shapes, a rectangular block 
of dimensions $9\, {\rm cm} \times 13\, {\rm cm} \times 7.6\, {\rm cm}$ 
(giving rise to integrable or possibly pseudo-integrable dynamics) and a 
regular tetrahedron with edges of length $10\, {\rm cm}$, where the 
dynamics is probably ergodic. In addition, we study each of the two bodies 
with and without an internal sphere. That sphere is suposed to provide another
free surface for the waves (rays) moving inside the body, which renders the 
dynamics chaotic. The inner sphere for the rectangle is chosen to be relatively
large in order to reduce bouncing ball orbits. The bodies (including the inner
spheres) are depicted in Fig.~\ref{f:geometries}. According to the random 
mode conversion model, the volume and the surface area of the bodies are 
important parameters. These are given in table~\ref{geotab}.

As explained earlier, the simulation
of the classical rays is done by launching a large number of rays (up to
$4\times 10^6$) with different initial conditions. Those are chosen to be 
either of type (ii) (Surface) or of type (iii) (Homogeneous). The former might 
be considered as corresponding to the experimental situation in 
Ref.~\cite{LobWea03,LobWea08}, but if at all this may be true only 
qualitatively.

\begin{table}
 \rowcolors{1}{white}{lightgray}
  \begin{tabular}{ccc}
  \toprule
   Body & Volume $[cm^3]$ & Surface $[cm^2]$\\
   \hline
   Rectangular block & $889.2$ & $568.4$ \\
   \ldots with inner sphere & $709.6$ & $722.3$ \\
   Regular tetrahedron & $117.8$ & $173.2$ \\
   \ldots with inner sphere & $110.6$ & $191.3 $ \\
  \toprule
  \end{tabular}
  \caption{}
\label{geotab}\end{table}

\subsection{Conversion rates and equipartition ratio}


\begin{figure}
\includegraphics[width=0.5\textwidth]{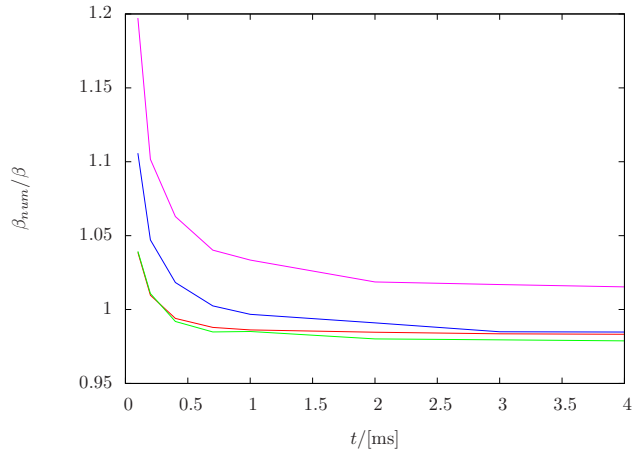}
\caption{The conversion rate $\beta_{\rm num}$ divided by the travel time $t$,
determined from the mean free $P$-mode occupation time, as a function of $t$,
for homogeneous initial conditions, for all four different geometries.
Tetrahedron with sphere (red line), without sphere (green line), Rectangle with 
sphere (blue line), and without sphere (pink line).}
\label{f:betastats}\end{figure}

To verify the accuracy of the random conversion model outlined above, we start
by analysing the conversion rates. For that purpose we perform simulations 
where the initial conditions of the rays are chosen according to the expected
equilibrium state. Thus, we start rays at random positions inside the body,
with random directions, in $S$- or $P$-mode according to the theoretical
equipartition ratio~(\ref{theoR}), and if in $S$-mode we choose the
polarization direction also at random. For each ray of pre-defined duration
$t$, we then record all periods during which the ray happened to travel in
$P$-mode. The average over those periods over all rays is just the mean free
$P$-mode travel time, and therefore equal to $\beta^{-1}$. In
Fig.~\ref{f:betastats} the so determined conversion rate $\beta_{\rm num}$ is 
compared to the theoretical estimate~(\ref{betatheo}) for all four geometries 
considered. We find that as soon as $t$ is large enough (for the random mode
conversion model to become valid, there need to occur sufficiently many 
reflections), $\beta_{\rm num}$ is quite close to the theoretical estimate. 
In fact, as can be observed in Fig.~\ref{f:betastats}, $\beta_{\rm num}$ never
deviates more than $3\%$ from the theoretical estimate. Nevertheless, there
are systematic differences for the different geometries. While 
$\beta_{\rm num}$ ends up about $2\%$ above the theoretical estimate in the
case of the rectangular block with inner sphere, $\beta_{\rm num}$ ends up 
about $3\%$ below, in the other three cases. We believe that this 
behavior is still acceptable in view of the fact that Eq.~(\ref{betatheo})
relies on rather rough estimates.

\subsubsection{Equipartition ratio}

\begin{figure}
\includegraphics[width=0.5\textwidth]{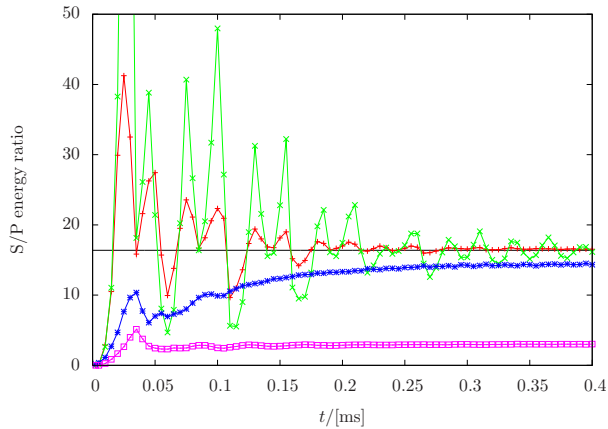}
\caption{S-mode vs. P-mode energy ratio for samples of different geometries.
Rays are started on the surface in $P$-mode with random directions (Surface 
type initial conditions). The color coding for the different geometries is the
same as in Fig.~\ref{f:betastats}. The black horizontal line shows the 
equipartition ratio $R$, according to Eq.~(\ref{theoR}).}
\label{f:StoP}\end{figure}

\begin{figure}
\includegraphics[width=0.5\textwidth]{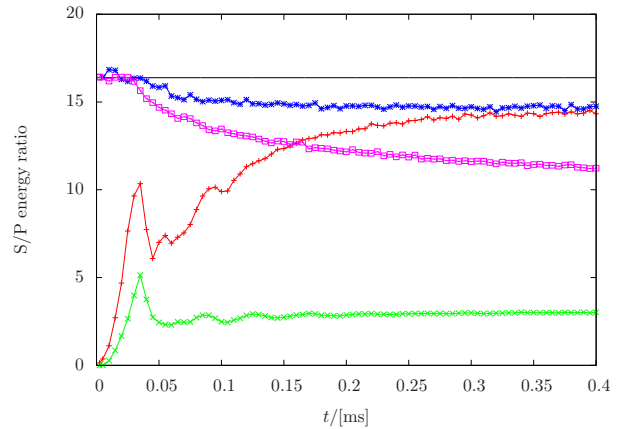}
\caption{S-mode vs. P-mode energy ratio for the rectangular block with and 
 without internal sphere. The initial conditions were chosen homogeneous and 
 isotropic (blue and pink line) and starting from a fixed point on the 
 surface (always starting in $P$-mode) (red and green line). The black 
 horizontal line shows the value of the equipartition ratio $R$, 
 Eq.~(\ref{theoR}).}
\label{f:StoP-EP}\end{figure}

In the Figs.~\ref{f:StoP} and~\ref{f:StoP-EP}, we study the energy share
between $S$- and $P$-mode rays. Since our simulations assume an equal amount of
energy associated with each ray, that energy share can be computed from the
relative frequencies with which we find a particular realization of the rays 
in either one of the two possible modes. In Figs.~\ref{f:StoP} 
and~\ref{f:StoP-EP}, we plot the ratio of these frequencies vs. $t$, the
travel time.

In Fig.~\ref{f:StoP} the simulation always starts with initial conditions on
the surface [initial conditions of type (ii)], where we applied the Monte 
Carlo sampling with $4\times 10^5$ random realizations, for each of the four
different geometries. Theoretically, we expect that the $S/P$ energy ratio
converges to $R= $ as given in Eq.~(\ref{theoR}). Since in the present case,
the rays always start in $P$-mode, the energy ratio at small times is close to
zero. We observe that two tetrahedrons show a similar behavior, which is very
distinct from that of the two rectangular blocks. They approach the 
theoretical value for $R$ via damped irregular oscillations which are 
initially very strong. Comparing the two tetrahedrons, the presence of the
inner sphere, tends to reduce the strength of the oscillations such that the
theoretical value for $R$ is reached faster. By contrast, the rectangular 
blocks show almost no oscillations at all, and quickly go over to a smooth 
approach of different limit values for the energy ratio. In doing so, the 
rectangle with inner sphere comes much closer to the theoretical value for $R$
than the rectangle without inner sphere.

Fig.~\ref{f:StoP-EP} shows simulations for the two rectangular blocks with
initial conditions of type (ii) (on the surface) and type (iii) (homogeneous). 
At small times, the energy ratio shown starts at zero for initial conditions 
on the surface, and at the value of the equipartition ratio $R$, for homogeneous 
initial conditions. As $t$ increases, the curves first show some minor 
fluctuations and then start to converge to different equilibrium values -- in 
all cases clearly below the equipartition ratio. For the rectangular block with
inner sphere, we observe that independent of the type of initial conditions,
the energy ratio converges to the same equilibrium value. For the rectangle 
without inner sphere, this is not the case, and the equilibrium values a very
different. 

Extending these simulations up to times of the order of $5\, {\rm ms}$, we have
confirmed that the $S/P$-energy ratio converges to finite values in the limit
of large times, except for the rectangular block without inner sphere. 
Surprisingly, we have found that these values do not depend on the type of 
initial conditions applied (Surface or Homogeneous). Concretely, we found the 
following values:
\begin{equation}
R= \left\{ \begin{array}{rcl}
 16.538 &:& \text{Tetrahedron without sphere}\\
 16.308 &:& \text{Tetrahedron with sphere}\\
 14.685 &:& \text{Rectangle with sphere}\\
  9.5/3.2 &:& \text{Rectangle without sphere}
\end{array} \right. \; ,
\label{numR}\end{equation}
while the theoretical value is $R= 16.384$.
We can see that the rectangular block without inner sphere must be considered
separately. Only in that case does the equilibrium distribution of the energy 
share depend on the initial state (for homogeneous initial conditions we get
$R=9.5$, otherwise $R=3.2$).

\subsection{\label{NP} P-mode occupation time statistics}

\begin{figure}
\includegraphics[width=0.5\textwidth]{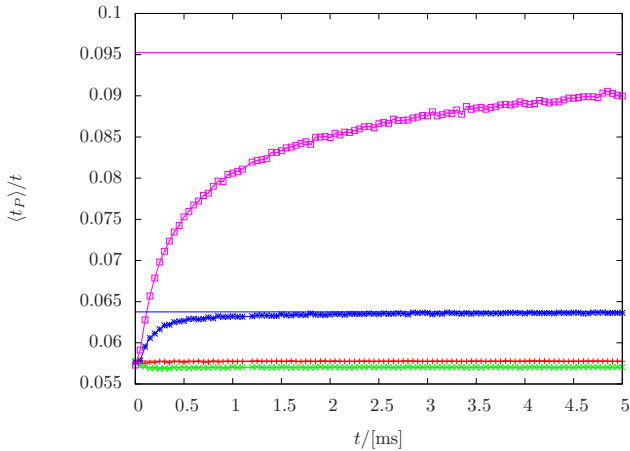}
\caption{For homogeneous initial conditions, the mean $P$-mode occupation time
 $\la t_P\ra$ for the different geometries, divided by the total travel 
 time $t$. We compare with the theoretical estimate, Eq.~(\ref{MeantpVartp}),
 where we have inserted the numerical values given in Eq.~(\ref{numR}).
The color coding is as before: Tetrahedron with inner sphere (red
 points), without inner sphere (green points), Rectangular block with inner 
 sphere (blue points), without inner sphere (pink points). The corresponding 
 estimates for the asymptotic value are depicted as horizontal lines of the 
 same color.}
\label{f:homogeneo2}\end{figure}

In this section, we restrict ourselves to homogeneous initial conditions.
In our simulations, every trajectory at each reflection makes a random choice 
whether to do a mode conversion or not. This results in different
trajectory paths and different amounts of times, the ray spends in each mode.
In what follows, we study the distribution of the $P$-mode occupation time
$t_P$; obviously, the corresponding $S$-mode occupation time would provide 
exactly the same information. We have chosen this particular quantity because
it may be linked to experimental results 
in~\cite{LobWea03,LobWea08} with the theoretical models proposed 
in~\cite{Sni06} (Coda wave interferometry) and~\cite{LobWea03} (random mode
conversion model).

We start with the average $P$-mode occupation time $\la t_P\ra$, for which the
random mode conversion model makes the prediction~(\ref{MeantpVartp}). In 
Fig.~\ref{f:tp-rcm}(a) we have seen that this prediction is indeed very 
accurate, provided the ensemble of rays is in the standard equilibrium state.
Fig.~\ref{f:homogeneo2} shows $\la t_P\ra/t$ for homogeneous initial conditions
which according to theory should converge to $(1+R)^{-1}$ at sufficiently large
times. Indeed for the tetrahedron as well as the rectangle with inner sphere,
the prediction is fulfilled. The ratio $\la t_P\ra/t$ converges to the 
predicted value, if we replace the theoretical equipartition ratio with the one
obtained numerically from our simulations (these values are given in 
Eq.~(\ref{numR})). In the case of the rectangle without inner sphere (out of
range), the convergence of $\la t_P\ra/t$ is much slower but the limit value
still consistent with the numerical $S/P$ energy partitioning.

\begin{figure}
\includegraphics[width=0.5\textwidth]{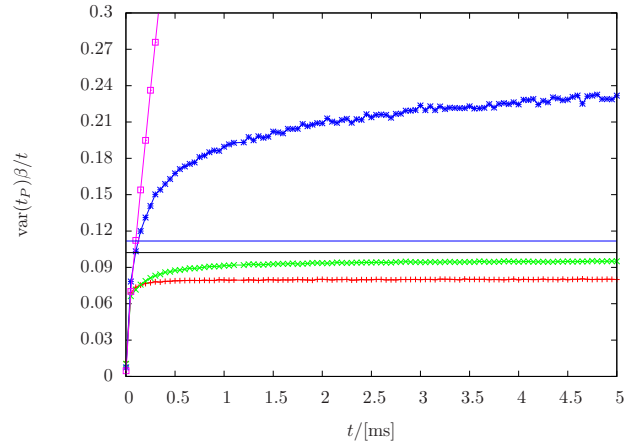}
\caption{For homogeneous initial conditions, the variance of the $P$-mode 
occupation times for the different geometries, rescaled by $\beta/t$. The 
color coding for the different geometries is the same as in 
Fig.~\ref{f:homogeneo2}.
We compare with the theoretical estimate, Eq.~(\ref{MeantpVartp}), where
$\beta$ has dropped out (black horizontal line). Correcting the theoretical
expectation for the $\beta$-values observed in Fig.~\ref{f:betastats}, leads 
to the blue horizontal line for the rectangular block with inner sphere.}
\label{f:homogeneo4}\end{figure}

Fig.~\ref{f:homogeneo4} shows the variance of the $P$-mode occupation times,
divided by $t/\beta$ with $\beta$ calculated from Eq.~(\ref{betatheo}). 
The black solid horizontal line shows the theoretically expected value 
according to the random mode conversion model~(\ref{MeantpVartp}), where 
$\beta$ drops out (black horizontal line):
\begin{equation}
{\rm var}(t_P)\; \frac{\beta}{t} \;\to\;
  \frac{2R^2}{(1+R)^3} \approx 0.1022 \; .
\label{theovartp}\end{equation}
Correcting the theoretical 
expectation for the numerically determined $\beta$-values shown in 
Fig.~\ref{f:betastats} does not make a noticeable difference in the case of 
the Tetrahedrons and leads to the blue horizontal line for the rectangular 
block with inner sphere. In the case of the rectangular block without sphere,
we find a steep linear increase of ${\rm var}(t_P)\, \beta/t$, such that a 
comparison with a theoretical limit value makes no sense. 

In this last figure we find the largest deviations from
the random mode conversion model. The result for the tetrahedron with sphere
is about $25\%$ below the theoretical value; without sphere, it is about $5\%$
below, while the result for the rectangle with sphere is more than $100\%$
above the theoretical value. 
The result for the rectangle without sphere is totally off the scale.

\begin{figure}
\includegraphics[width=0.5\textwidth]{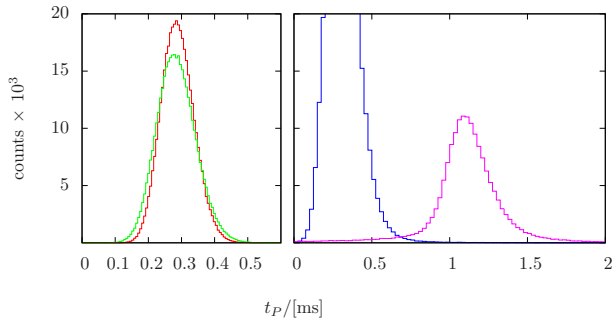}
\caption{For homogeneous initial conditions, the histograms for the $t_P$
statistics for $t= 5\, {\rm ms}$. The result for the tetrahedrons on the left,
with/without inner sphere (red/green solid line); for the rectangular blocks
on the right with/without inner sphere (blue/pink solid lines).}
\label{f:tp-hists}\end{figure}

Finally, we show in Fig.~\ref{f:tp-hists} the distribution of $P$-mode 
occupation times $t_P$ for trajectories of duration $t= 5\, {\rm ms}$. In the
case of the tetrahedrons, we find that the distributions are close to 
Gaussians, with the distribution for the tetrahedron with inner sphere being
a bit narrower than for the tetrahedron without. In the case of the rectangular
blocks, the distributions are clearly non-Gaussian. For the rectangle with
inner sphere the distribution is strongly asymmetric, while for the rectangle
without inner sphere, the distributions is surprisingly close to a Lorentzian.
This latter observation can explain the fact that the variance of the $P$-mode
occupation times scales with $t^2$ rather than $t$, since the variance of a 
Lorentzian distribution is infinite.

\section{\label{D} Discussion}

\subsection{\label{DE} Related experimental results}

In Refs.~\cite{LobWea03,LobWea08}, the authors measured the acoustic long time
response of short initial ultrasound pulses applied to different Aluminum 
bodies. They studied 
the cross correlations between these signals taken at different temperatures. 
Since the temperature change induces a change of the propagation speeds of
$P$- and $S$-waves by different amounts, a temperature change results in the 
distortion of the acoustic signal and a reduction of the cross correlations. The 
reduction of the cross correlations, which has also been identified as a 
scattering fidelity~\cite{SGSS05,SSGS05,GPSZ06}, can be described 
quantitatively within the theory of Coda wave interferometry~\cite{Sni06}. 

As this theory shows, scattering fidelity (or ``distortion'' how that quantity 
has been called in 
Ref.~\cite{LobWea03}) is essentially given by the distribution of $P$-mode
occupation times -- the quantity studied in the previous Sec.~\ref{NP}. 
Assuming Gaussian statistics for these times, the scattering fidelity or 
distortion may be related to the variance of the $P$-mode occupation times
as follows:
\begin{equation}
D(t) = 2\pi^2\; (\Delta T\, f)^2\; (\delta_p - \delta_S)^2\; 
  {\rm var}(t_p) \; ,
\end{equation}
where $f$ is the carrier frequency of the elastic wave, $\Delta T$ is the 
change in temperature, while $\delta_P = -1.685\times 10^{-4}\, {\rm K}^{-1}$,
and $\delta_S = -2.9\times 10^{-4}\, {\rm K}^{-1}$ are the thermal dilation
coefficients for $P$-waves and $S$-waves, respectively~\cite{LobWea03}. 

In Ref.~\cite{LobWea03}, Lobkis and Weaver measured the slope of $D(t)$ scaled 
by $\Delta T^2$, $f^2$ and $\beta^{-1}$, which according to the random mode
conversion model should always be the same. For our simulations, this is 
equivalent to determining the slope of ${\rm var}(t_P)$ as a function of time,
scaled by $\beta^{-1}$, which should then also give a unique value; see 
Eq.~(\ref{theovartp}). For the Aluminum blocks of different shapes, analysed in 
Ref.~\cite{LobWea03}, the result was a wide spreading of experimental values 
where, the smallest values roughly agreed with the theoretical expectation, 
while the largest values where about four times larger
\footnote{Ignoring the case of a regular cylinder, which has been off the 
scale.}.
On a qualitative basis, one could observe the tendency that less 
chaotic geometries lead to larger values for the slope. 


\subsection{\label{DO} Our results}

The overall result of the present analysis, shown in Fig.~\ref{f:homogeneo4}, 
is similar in this respect. We also find that the slopes can be very different
for different geometries. It however shows that the relevant quantity is not
really chaos as measured by Lyapunov exponents but possibly rather ergodicity. 
From the 
four different geometries considered, the tetrahedron with inner sphere has 
the smallest slope. It is notable in that case, that the slope is about $25\%$
below the theoretical value, which shows that the theory doesn't provide a lower
bound as one might have been conjectured from the experimental results. 
Next comes the tetrahedron without inner sphere. Because 
all its surfaces are plane, the Lyapunov exponent must be zero in any case. 
However, the tetrahedron has most likely ergodic dynamics. And we find indeed
that the slope is quite close to the theoretical expectation. Only at a 
considerable distance, we find the rectangle with inner sphere. The inner 
sphere clearly leads to chaotic dynamics with positive Lyapunov exponents, but
there are also regions of integrable dynamics and bouncing ball orbits. That
is apparently sufficient to increase the slope to more than twice the 
theoretical value. The rectangle without inner sphere is clearly the most 
regular body. For that geometry, the ${\rm var}(t_P)$ curve rather shows a 
quadratic dependence on $t$. This is in line with the probability density for
the $P$-mode occupation times shown in Fig.~\ref{f:tp-hists}. Its shape is 
almost a Lorentzian which would imply an infinite variance. 

Our simulations have shown that many of the assumptions within the random 
mode conversion model are rather nicely met. This is true for the $P$-to-$S$
conversion rate $\beta$ (Fig.~\ref{f:betastats}), and also for the 
equipartition ratio if the dynamics is ergodic (Fig.~\ref{f:StoP}). We could 
also confirm that the average $P$-mode occupation time is always related to 
the $S/P$-energy ratio (Fig.~\ref{f:homogeneo2}). Thus, the weak point of the
random mode conversion model is clearly the in general inaccurate or wrong 
prediction of the variance of the $P$-mode occupation times. This shows that
the succession of $P$-mode and $S$-mode segments is in general not well 
described by a Poissonian process. We believe that there are two effects 
coming into play. (i) The duration of the individual $P$-mode segments (these
are the shortest ones) cannot be really exponentially distributed because the
rays have to travel a certain distance before having the possibility to undergo
a mode conversion. 
(ii) The durations of subsequent $P$-mode and $S$-mode segments might be 
correlated.

The first effect would lead to a smaller variance of the duration of the
$P$-mode segments and thereby to a smaller variance of the total $P$-mode
occupation times. Thus if the dynamics destroys correlations sufficiently 
rapidly, ${\rm var}(t_P)$ should be smaller than expected theoretically. Our
findings suggest that this is indeed the case for the tetrahedron with inner
sphere. The tetrahedron without inner sphere is ergodic but needs more time to
destroy correlations. 
It thus seems that the second effect 
related to correlations tends to enlarge ${\rm var}(t_P)$.
For the rectangle with inner sphere, correlations are not efficiently 
destroyed and the variance of the $P$-mode occupation times becomes much larger 
still. At last, we have the rectangle without inner sphere, where 
${\rm var}(t_P)$ scales with $t^2$.

\section{\label{C} Conclusions}

We presented simulations of the propagation of elastic waves in 
three-dimensional bodies of different geometries in the limit of classical
rays. Because of mode conversion at the reflections on the surface of the 
bodies, one has to deal with an exponential proliferation of branches of the
elastic ray, which is dealt with using Monte Carlo sampling. 

Our simulations have shown that there is no unique universal equilibrium 
distribution for elastic rays, at least not for bodies with a sufficiently 
simple geometry.
This is true even if the dynamics must be considered as completely chaotic. For
the tetrahedron with internal sphere, the equilibrium limit of the $S/P$ 
energy ratio was close but not exactly equal to the theoretical value. Even 
more surprisingly, we found that the homogeneous and isotropic distribution of
elastic rays with an $S/P$ energy ratio equal to the theoretical equipartition
ratio need not be an equilibrium distribution at all. When chosen as the 
initial condition of an ensemble of elastic rays, we found that in the case of
the rectangular blocks, the $S/P$ energy ratio changes to a different value. 

The main purpose of the present work was to analyse some of the
conjectures made in Ref.~\cite{LobWea03} and to investigate, whether the 
observed deviations can be explained with a model based on classical rays.
While many conjectures and
approximations were indeed justified, the most problematic assumption was that
of random mode conversions with given conversion rates. There, we found two 
counteracting effects: One the one hand the length distribution of individual
$P$-mode and $S$-mode segments is not exponential due to the fact that
mode conversions can take place only during reflections. On the other hand, 
there are correlations in time where the time scale depends on the dynamics. 
In comparison to the random mode conversion model, the former tends to reduce 
the variance of the $P$-mode occupation time while the latter leads to an 
increase. 

Previous publications~\cite{GSW06,LobWea08} have focused on the form of the 
decay function of scattering fidelity which has been shown to follow very
closely universal random matrix predictions. However, our results indicate that
the perturbation strength is an equally interesting quantity. In fact it 
reveals much more system specific information, which may even have practical
applications, for example in the analysis and verification of the properties 
of mechanical components. It would be interesting to design new experiments,
which allowed for a quantitative comparison between experiment and numerical 
simulations.

\begin{acknowledgments}
We are grateful to discussions with O.~I. Lobkis and R.~L. Weaver at an early
stage of the project. We are also grateful to the hospitality of the Centro
Internacional de Ciencias (Cuernavaca, M\' exico) where this discussions took 
place. Finally, we acknowledge the support by CONACyT under grant 
number 129309.
\end{acknowledgments}

\begin{appendix}

\section{Energy conservation during ray splitting}

In Sec.~\ref{RC}, we discuss the splitting of the total energy of a reflected
wavepacket in the ray limit. In order to determine the energy
share of the mode-converted branch, we assumed energy conservation, which
is an important issue for the self-consistency of our ray model.
It implies that the reflection coefficients defined in 
Eqs.~(\ref{RPP}-\ref{RSSp}) are related. Here, we show exemplarily for an 
incident ray in $P$-mode that the total energy is indeed conserved.

\begin{figure}[t]
\includegraphics[width=0.5\textwidth]{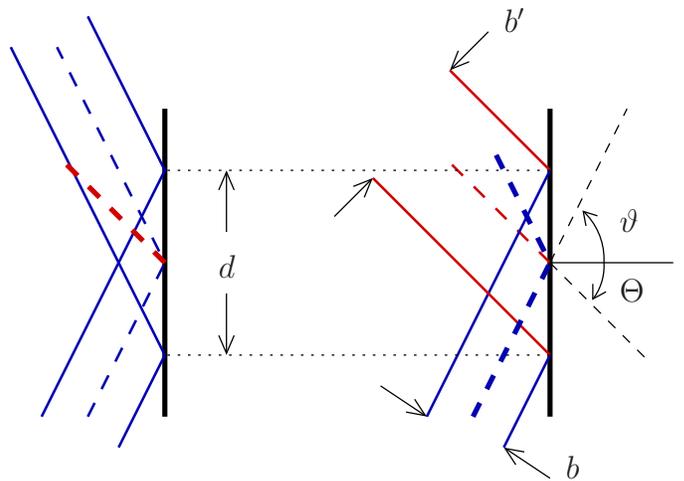}
\caption{Transformation of the ray widths under mode conversion. If the exit
angle is equal to the angle of incidence (no mode conversion), the width of the
reflected ray is unchanged. Otherwise the widths are proportional to the
cosines of the angles $\vartheta$ and $\Theta$.}
\label{f:raywidths}\end{figure}

The amplitude of an elastic ray is physically related to the displacement
of infinitesimal volume elements in the solid. Therefore, amplitudes $u_P$ or
$u_S$ (for $P$-mode and $S$-mode, respectively) have units of length. The 
period averaged energy density of the respective wave field is then given 
by 
\begin{equation}
\eps_{P,S} = \frac{\varrho\, w^2}{2}\; u_{\{P,S\}}^2 \; ,    
\end{equation}                                             
where $\varrho$ is the density of the medium and
$w$ the angular frequency of the wave 
(see Sec.~1.7 of~\cite{Ach73}). To demonstrate the energy
conservation, we show that the energy flux is conserved in a 
quasi-stationary situation where a very long wave packet is 
reflected at a free plane surface.
It means that the energy flux along the propagation direction of the incident 
$P$-mode ray must be equal to the sum of the fluxes along the reflected 
$P$-mode and the reflected and mode-converted $S$-mode branch: In general, the 
amount of energy flowing through a transversal surface $S$ is given by
\begin{equation}
F = \int_S\eps(\vec r)\; v_n(\vec r)\; \rmd\sigma(\vec r) \; ,
\end{equation}
where $v_n$ is the projection of the ray velocity on the surface normal. If we 
choose $S$ to be normal to the propagation direction:
\begin{equation}
F_{P/S} = \frac{\varrho\, w^2}{2}\; c_{d/s} 
   \int_S u_{P/S}(\vec r)^2\; \rmd\sigma(\vec r) \; . 
\end{equation}
Just as the propagation speed, also the energy flux depends on the wave mode. 
Now, if the energy flux is really conserved, the following relation must hold:
\begin{eqnarray}
&& \frac{\varrho w^2}{2}\; c_d\int_S u_P(\vec r)^2\; \rmd\sigma(\vec r)
 = \frac{\varrho w^2}{2}\nonumber\\
&&\times \left[ 
      c_d\; R_{PP}^2\int_{S'} u_P'(\vec r)^2\, \rmd\sigma(\vec r)
    + c_s\; R_{SP}^2\int_{S''} u_P''(\vec r)^2\, \rmd\sigma(\vec r) \right]
 \nonumber\\
&&\quad
\end{eqnarray}
where $S$ is a surface perpendicular to the incident ray, $S'$ is a surface
perpendicular to the reflected $P$-mode branch and $S''$ is a surface 
perpendicular to the reflected $S$-mode branch. We have assumed that these 
surfaces are sufficiently to the reflection point such that the wave amplitudes
may be considered constant along the line segments towards and away from the
reflection point. The reflection is schematically depicted in 
Fig.~\ref{f:raywidths} separately for the $P$-mode branch (left hand side) and 
the $S$-mode branch (right hand side). Without mode conversion, the geometrical
properties of the ray do not change, such that the amplitude $u_P'$ relative to
$S'$ is just equal to $u_P$ relative to $S$, which means that the corresponding
integrals coincide. With mode conversion, the width of the ray increases by the
factor $\cos\Theta/\cos\vartheta > 1$. Thus, the integral over $S''$ must be
scaled by that factor. Therefore, dividing by the common factor 
$\varrho w^2/2$,
\begin{eqnarray}
c_d\; (1 - R_{PP}^2)&&\int_S u_P(\vec r)^2\; \rmd\sigma(\vec r) = 
  c_s\; R_{SP}^2\; \frac{\cos\Theta}{\cos\vartheta} \nonumber\\
&&\qquad\times \int_S u_P(\vec r)^2\; \rmd\sigma(\vec r)
\end{eqnarray}
Now, since the integrals are the same, we arrive at
\begin{equation}
1 - R_{PP}^2 = \frac{c_s}{c_d}\; R_{SP}^2\; \frac{\cos\Theta}{\cos\vartheta}
\; ,
\end{equation}
which may be easily verified. Thus the energy flux is indeed conserved.

\end{appendix}

\bibliography{JabRef}

\begin{thebibliography}{10}%
\makeatletter
\providecommand \@ifxundefined [1]{%
 \ifx #1\undefined \expandafter \@firstoftwo
 \else \expandafter \@secondoftwo
\fi
}%
\providecommand \@ifnum [1]{%
 \ifnum #1\expandafter \@firstoftwo
 \else \expandafter \@secondoftwo
\fi
}%
\providecommand \enquote [1]{``#1''}%
\providecommand \bibnamefont  [1]{#1}%
\providecommand \bibfnamefont [1]{#1}%
\providecommand \citenamefont [1]{#1}%
\providecommand\href[0]{\@sanitize\@href}%
\providecommand\@href[1]{\endgroup\@@startlink{#1}\endgroup\@@href}%
\providecommand\@@href[1]{#1\@@endlink}%
\providecommand \@sanitize [0]{\begingroup\catcode`\&12\catcode`\#12\relax}%
\@ifxundefined \pdfoutput {\@firstoftwo}{%
 \@ifnum{\z@=\pdfoutput}{\@firstoftwo}{\@secondoftwo}%
}{%
 \providecommand\@@startlink[1]{\leavevmode\special{html:<a href="#1">}}%
 \providecommand\@@endlink[0]{\special{html:</a>}}%
}{%
 \providecommand\@@startlink[1]{%
  \leavevmode
  \pdfstartlink
   attr{/Border[0 0 1 ]/H/I/C[0 1 1]}%
   user{/Subtype/Link/A<</Type/Action/S/URI/URI(#1)>>}%
  \relax
 }%
 \providecommand\@@endlink[0]{\pdfendlink}%
}%
\providecommand \url  [0]{\begingroup\@sanitize \@url }%
\providecommand \@url [1]{\endgroup\@href {#1}{\urlprefix}}%
\providecommand \urlprefix [0]{URL }%
\providecommand \Eprint[0]{\href }%
\@ifxundefined \urlstyle {%
  \providecommand \doi [1]{doi:\discretionary{}{}{}#1}%
}{%
  \providecommand \doi [0]{doi:\discretionary{}{}{}\begingroup
  \urlstyle{rm}\Url }%
}%
\providecommand \doibase [0]{http://dx.doi.org/}%
\providecommand \Doi[1]{\href{\doibase#1}}%
\providecommand \bibAnnote [3]{%
  \BibitemShut{#1}%
  \begin{quotation}\noindent
    \textsc{Key:}\ #2\\\textsc{Annotation:}\ #3%
  \end{quotation}%
}%
\providecommand \bibAnnoteFile [2]{%
  \IfFileExists{#2}{\bibAnnote {#1} {#2} {\input{#2}}}{}%
}%
\providecommand \typeout [0]{\immediate \write \m@ne }%
\providecommand \selectlanguage [0]{\@gobble}%
\providecommand \bibinfo [0]{\@secondoftwo}%
\providecommand \bibfield [0]{\@secondoftwo}%
\providecommand \translation [1]{[#1]}%
\providecommand \BibitemOpen[0]{}%
\providecommand \bibitemStop [0]{}%
\providecommand \bibitemNoStop [0]{.\EOS\space}%
\providecommand \EOS [0]{\spacefactor3000\relax}%
\providecommand \BibitemShut [1]{\csname bibitem#1\endcsname}%
\bibitem{Sni06}%
  \BibitemOpen
  \bibfield{author}{%
  \bibinfo {author} {\bibfnamefont{R.}~\bibnamefont{Snieder}},\ }%
  \bibfield{journal}{%
  \bibinfo {journal} {Pure appl. geophys.}\ }%
  \textbf{\bibinfo {volume} {163}},\ \bibinfo {pages} {455} 
   (\bibinfo {year} {2006})%
   \bibAnnoteFile{NoStop}{Sni06}
\bibitem{Courtland08}%
  \BibitemOpen
  \bibfield{author}{%
  \bibinfo {author} {\bibfnamefont{R.}~\bibnamefont{Courtland}},\ }%
  \bibfield{journal}{%
  \bibinfo {journal} {Nature}\ }%
  \textbf{\bibinfo {volume} {453}},\ \bibinfo {pages} {146}
  (\bibinfo {year} {2008})%
  \bibAnnoteFile{NoStop}{Courtland08}%
\bibitem{Larose06}%
  \BibitemOpen
  \bibfield{author}{%
  \bibinfo {author} {\bibfnamefont{E.}~\bibnamefont{Larose}}, \bibinfo {author}
  {\bibfnamefont{J.}~\bibnamefont{De~Rosny}}, \bibinfo {author}
  {\bibfnamefont{L.}~\bibnamefont{Margerin}}, \bibinfo {author}
  {\bibfnamefont{D.}~\bibnamefont{Anache}}, \bibinfo {author}
  {\bibfnamefont{P.}~\bibnamefont{Gouedard}}, \bibinfo {author}
  {\bibfnamefont{M.}~\bibnamefont{Campillo}},\ and\ \bibinfo {author}
  {\bibfnamefont{B.}~\bibnamefont{van Tiggelen}},\ }%
  \bibfield{journal}{%
  \Doi{10.1103/PhysRevE.73.016609}
  {\bibinfo {journal} {Phys. Rev. E}}\ }%
  \textbf{\bibinfo {volume} {73}},\ \bibinfo {pages} {016609} 
  (\bibinfo {year} {2006})%
  \bibAnnoteFile{NoStop}{Larose06}%
\bibitem{KohBlu98}%
  \BibitemOpen
  \bibfield{author}{%
  \bibinfo {author} {\bibfnamefont{A.}~\bibnamefont{Kohler}}\ and\ \bibinfo
  {author} {\bibfnamefont{R.}~\bibnamefont{Bl\" umel}},\ }%
  \bibfield{journal}{%
  \bibinfo {journal} {Ann. Phys.}\ }%
  \textbf{\bibinfo {volume} {267}},\ \bibinfo {pages} {249} 
  (\bibinfo {year} {1998})
  \bibAnnoteFile{NoStop}{KohBlu98}%
\bibitem{Berry85}%
  \BibitemOpen
  \bibfield{author}{%
  \bibinfo {author} {\bibfnamefont{M.~V.}\ \bibnamefont{Berry}},\ }%
  \bibfield{journal}{%
  \bibinfo {journal} {Proc. R. Soc. Lond. A}\ }%
  \textbf{\bibinfo {volume} {400}},\ \bibinfo {pages} {229} (\bibinfo {year}
  {1985})%
  \bibAnnoteFile{NoStop}{Berry85}%
\bibitem{JacBee01}%
  \BibitemOpen
  \bibfield{author}{%
  \bibinfo {author} {\bibfnamefont{P.}~\bibnamefont{Jacquod}}, \bibinfo
  {author} {\bibfnamefont{P.~G.}\ \bibnamefont{Silvestrov}},\ and\ \bibinfo
  {author} {\bibfnamefont{C.~W.~J.}\ \bibnamefont{Beenakker}},\ }%
  \bibfield{journal}{%
  \bibinfo {journal} {Phys. Rev. E}\ }%
  \textbf{\bibinfo {volume} {64}},\ \bibinfo {pages} {055203(R)} 
  (\bibinfo {year} {2001})%
  \bibAnnoteFile{NoStop}{JacBee01}%
\bibitem{CerTom02}%
  \BibitemOpen
  \bibfield{author}{%
  \bibinfo {author} {\bibfnamefont{N.~R.}\ \bibnamefont{Cerruti}}\ and\
  \bibinfo {author} {\bibfnamefont{S.}~\bibnamefont{Tomsovic}},\ }%
  \bibfield{journal}{%
  \bibinfo {journal} {Phys. Rev. Lett.}\ }%
  \textbf{\bibinfo {volume} {88}},\ \bibinfo {pages} {054103} 
  (\bibinfo {year} {2002})%
  \bibAnnoteFile{NoStop}{CerTom02}%
\bibitem{LobWea03}%
  \BibitemOpen
  \bibfield{author}{%
  \bibinfo {author} {\bibfnamefont{O.~I.}\ \bibnamefont{Lobkis}}\ and\ \bibinfo
  {author} {\bibfnamefont{R.~L.}\ \bibnamefont{Weaver}},\ }%
  \bibfield{journal}{%
  \bibinfo {journal} {Phys. Rev. Lett.}\ }%
  \textbf{\bibinfo {volume} {90}},\ \bibinfo {pages} {254302} 
  (\bibinfo {year} {2003})%
  \bibAnnoteFile{NoStop}{LobWea03}%
\bibitem{SGSS05}%
  \BibitemOpen
  \bibfield{author}{%
  \bibinfo {author} {\bibfnamefont{R.}~\bibnamefont{Sch\" afer}}, \bibinfo
  {author} {\bibfnamefont{T.}~\bibnamefont{Gorin}}, \bibinfo {author}
  {\bibfnamefont{H.-J.}\ \bibnamefont{St\" ockmann}},\ and\ \bibinfo {author}
  {\bibfnamefont{T.~H.}\ \bibnamefont{Seligman}},\ }%
  \bibfield{journal}{%
  \bibinfo {journal} {New J. Phys.}\ }%
  \textbf{\bibinfo {volume} {7}},\ \bibinfo {pages} {152} 
  (\bibinfo {year} {2005})%
  \bibAnnoteFile{NoStop}{SGSS05}%
\bibitem{SSGS05}%
  \BibitemOpen
  \bibfield{author}{%
  \bibinfo {author} {\bibfnamefont{R.}~\bibnamefont{Sch\" afer}}, \bibinfo
  {author} {\bibfnamefont{H.-J.}\ \bibnamefont{St\" ockmann}}, 
  \bibinfo {author} {\bibfnamefont{T.}~\bibnamefont{Gorin}},\ and\ 
  \bibinfo {author} {\bibfnamefont{T.~H.}\ \bibnamefont{Seligman}},\ }%
  \bibfield{journal}{%
  \bibinfo {journal} {Phys. Rev. Lett.}\ }%
  \textbf{\bibinfo {volume} {95}},\ \bibinfo {pages} {184102} 
  (\bibinfo {year} {2005})%
  \bibAnnoteFile{NoStop}{SSGS05}%
\bibitem{GSW06}%
  \BibitemOpen
  \bibfield{author}{%
  \bibinfo {author} {\bibfnamefont{T.}~\bibnamefont{Gorin}}, \bibinfo {author}
  {\bibfnamefont{T.~H.}\ \bibnamefont{Seligman}},\ and\ \bibinfo {author}
  {\bibfnamefont{R.~L.}\ \bibnamefont{Weaver}},\ }%
  \bibfield{journal}{%
  \bibinfo {journal} {Phys. Rev. E}\ }%
  \textbf{\bibinfo {volume} {73}},\ \bibinfo {pages} {015202(R)} 
  (\bibinfo {year} {2006})%
  \bibAnnoteFile{NoStop}{GSW06}%
\bibitem{GPSZ06}%
  \BibitemOpen
  \bibfield{author}{%
  \bibinfo {author} {\bibfnamefont{T.}~\bibnamefont{Gorin}}, \bibinfo {author}
  {\bibfnamefont{T.}~\bibnamefont{Prosen}}, \bibinfo {author}
  {\bibfnamefont{T.~H.}\ \bibnamefont{Seligman}},\ and\ \bibinfo {author}
  {\bibfnamefont{M.}~\bibnamefont{\v{Z}nidari\v{c}}},\ }%
  \bibfield{journal}{%
  \bibinfo {journal} {Phys. Rep.}\ }%
  \textbf{\bibinfo {volume} {435}},\ \bibinfo {pages} {33} 
  (\bibinfo {year} {2006})%
  \bibAnnoteFile{NoStop}{GPSZ06}%
\bibitem{LobWea08}%
  \BibitemOpen
  \bibfield{author}{%
  \bibinfo {author} {\bibfnamefont{O.~I.}\ \bibnamefont{Lobkis}}\ and\ \bibinfo
  {author} {\bibfnamefont{R.~L.}\ \bibnamefont{Weaver}},\ }%
  \bibfield{journal}{%
  \Doi{10.1103/PhysRevE.78.066212}{\bibinfo {journal} {Phys. Rev. E}}\ }%
  \textbf{\bibinfo {volume} {78}},\ \bibinfo {eid} {066212} (\bibinfo {year}
  {2008})%
  \bibAnnoteFile{NoStop}{LobWea08}%
\bibitem{Wea82}%
  \BibitemOpen
  \bibfield{author}{%
  \bibinfo {author} {\bibfnamefont{R.~L.}\ \bibnamefont{Weaver}},\ }%
  \bibfield{journal}{%
  \bibinfo {journal} {J. Acoust. Soc. Am.}\ }%
  \textbf{\bibinfo {volume} {71}},\ \bibinfo {pages} {1608} 
  (\bibinfo {year} {1982})%
  \bibAnnoteFile{NoStop}{Wea82}%
\bibitem{Pierce1981}%
  \BibitemOpen
  \bibfield{author}{%
  \bibinfo {author} {\bibfnamefont{A.~D.}\ \bibnamefont{Pierce}},\ }%
  \emph{\bibinfo {title} {Acoustics: An Introduction}}\ (\bibinfo
  {publisher} {McGraw-Hill},\ \bibinfo {year} {1981})%
  \bibAnnoteFile{NoStop}{Pierce1981}%
\bibitem{Graff1991}%
  \BibitemOpen
  \bibfield{author}{%
  \bibinfo {author} {\bibfnamefont{K.~F.}\ \bibnamefont{Graff}},\ }%
  \emph{\bibinfo {title} {Wave Motion in Elastic Solids}}\ (\bibinfo
  {publisher} {Dover},\ \bibinfo {year} {1991})%
  \bibAnnoteFile{NoStop}{Graff1991}%
\bibitem{SchoJac05}%
  \BibitemOpen
  \bibfield{author}{%
  \bibinfo {author} {\bibfnamefont{H.}\ \bibnamefont{Schomerus}}\ and\ \bibinfo
  {author} {\bibfnamefont{P.}\ \bibnamefont{Jacquod}},\ }%
  \bibfield{journal}{%
  {\bibinfo {journal} {J. Phys. A: Math. Gen.}}\ }%
  \textbf{\bibinfo {volume} {38}},\ \bibinfo {eid} {10663} (\bibinfo {year}
  {2005})%
  \bibAnnoteFile{NoStop}{LobWea08}%
\bibitem{Gutzwiller90}%
  \BibitemOpen
  \bibfield{author}{%
  \bibinfo {author} {\bibfnamefont{M.~C.}\ \bibnamefont{Gutzwiller}},\ }%
  \emph{\bibinfo {title} {Chaos in classical and quantum mechanics}}\ (\bibinfo
  {publisher} {Springer},\ \bibinfo {year} {1990})%
  \bibAnnoteFile{NoStop}{Gutzwiller90}%
\bibitem{Kap02}%
  \BibitemOpen
  \bibfield{author}{%
  \bibinfo {author} {\bibfnamefont{L.}~\bibnamefont{Kaplan}},\ }%
  \bibfield{journal}{%
  \bibinfo {journal} {New J. Phys.}\ }%
  \textbf{\bibinfo {volume} {4}},\ \bibinfo {pages} {90} 
  (\bibinfo {year} {2002})%
  \bibAnnoteFile{NoStop}{Kap02}%
\bibitem{BPS00}%
  \BibitemOpen
  \bibfield{author}{%
  \bibinfo {author} {\bibfnamefont{E.}~\bibnamefont{Bogomolny}}, 
  \bibinfo {author} {\bibfnamefont{N.}~\bibnamefont{Pavloff}}\ and\ 
  \bibinfo {author} {\bibfnamefont{C.}~\bibnamefont{Schmit}},\ }%
  \bibfield{journal}{%
  \bibinfo {journal} {Phys. Rev. E}\ }%
  \textbf{\bibinfo {volume} {61}},\ \bibinfo {pages} {3689} 
  (\bibinfo {year} {2000})%
  \bibAnnoteFile{NoStop}{BPS00}%
\bibitem{HHH00}%
  \BibitemOpen
  \bibfield{author}{%
  \bibinfo {author} {\bibfnamefont{J.~S.}~\bibnamefont{Hersch}}, 
  \bibinfo {author} {\bibfnamefont{M.~R.}~\bibnamefont{Haggerty}}\ and\ 
  \bibinfo {author} {\bibfnamefont{E.~J.}~\bibnamefont{Heller}},\ }%
  \bibfield{journal}{%
  \bibinfo {journal} {Phys. Rev. E}\ }%
  \textbf{\bibinfo {volume} {62}},\ \bibinfo {pages} {4873} 
  (\bibinfo {year} {2000})%
  \bibAnnoteFile{NoStop}{BPS00}%
\bibitem{Ach73}%
  \BibitemOpen
  \bibfield{author}{%
  \bibinfo {author} {\bibfnamefont{J.~D.}\ \bibnamefont{Achenbach}},\ }%
  \emph{\bibinfo {title} {Wave propagation in elastic solids}},\ \bibinfo
  {series} {Applied mathematics and mechanics}, Vol.~\bibinfo {volume} {16}\
  (\bibinfo {publisher} {North-Holland},\ 
   \bibinfo {year} {1973})%
  \bibAnnoteFile{NoStop}{Ach73}%
\bibitem{Note1}%
  \BibitemOpen
  \bibinfo {note} {Ignoring the case of a regular cylinder, which has been
  off the scale.}%
  \bibAnnoteFile{Stop}{Note1}%
\end{thebibliography}%

\end{document}